\documentclass[reprint,prl]{revtex4-1}
\usepackage{amsmath,amssymb,amsfonts}
\usepackage{mathtools}
\usepackage{graphicx}
\usepackage{hyperref}
\usepackage{epstopdf}

\usepackage[capitalize]{cleveref}
\usepackage{bm}
\usepackage{footmisc}
\usepackage{float}

\begin{document}

\title{Multipartite entanglement generation and contextuality tests using non-destructive three-qubit parity measurements}

\author{S.B. van Dam} 
\affiliation{QuTech, Delft University of Technology, PO Box 5046, 2600 GA Delft, The Netherlands}
\affiliation{Kavli Institute of Nanoscience, Delft University of Technology, PO Box 5046, 2600 GA Delft, The Netherlands}

\author{J. Cramer}
\affiliation{QuTech, Delft University of Technology, PO Box 5046, 2600 GA Delft, The Netherlands}
\affiliation{Kavli Institute of Nanoscience, Delft University of Technology, PO Box 5046, 2600 GA Delft, The Netherlands}

\author{T.H. Taminiau}
\affiliation{QuTech, Delft University of Technology, PO Box 5046, 2600 GA Delft, The Netherlands}
\affiliation{Kavli Institute of Nanoscience, Delft University of Technology, PO Box 5046, 2600 GA Delft, The Netherlands}
 
\author{R. Hanson}
\email{r.hanson@tudelft.nl}
\affiliation{QuTech, Delft University of Technology, PO Box 5046, 2600 GA Delft, The Netherlands}
\affiliation{Kavli Institute of Nanoscience, Delft University of Technology, PO Box 5046, 2600 GA Delft, The Netherlands}

\begin{abstract}
We report on the realization and application of non-destructive three-qubit parity measurements on nuclear spin qubits in diamond. We use high-fidelity quantum logic to map the parity of the joint state of three nuclear spin qubits onto an electronic spin qubit that acts as an ancilla, followed by single-shot non-destructive readout of the ancilla combined with an electron spin echo to ensure outcome-independent evolution of the nuclear spins. Through the sequential application of three such parity measurements, we demonstrate the generation of genuine multipartite entangled states out of the maximally mixed state. Furthermore, we implement a single-shot version of the GHZ experiment that can generate a quantum versus classical contradiction in each run. Finally, we test a state-independent non-contextuality inequality in eight dimensions. The techniques and insights developed here are relevant for fundamental tests as well as for quantum information protocols such as quantum error correction.
\end{abstract}

\maketitle
Parity measurements - measurements that reveal whether the sum of a (quantum) bit string is even or odd - are a prime example of the radically different roles of measurement in quantum physics and classical physics. In contrast to classical parity measurements, a quantum parity measurement is able to extract only the parity information from the system without revealing any additional information about the individual qubit states. Therefore, the coherences within the parity subspace into which the system is projected remain unaffected. Thanks to these unique quantum properties parity measurements are at the heart of many quantum information protocols, for example as stabilizer measurements in quantum error correction \citep{Bravyi1998,Raussendorf2007} or to generate entangled states \citep{Barreiro2011,Riste2013,Pfaff2013,Saira2014,Cramer2015}. In addition, their strikingly non-classical behavior features in tests of the foundations of quantum mechanics \citep{Bell1966,Kochen1967}.

Experimentally, realizing parity measurements that project a system on a parity subspace but are otherwise non-destructive is challenging: uncontrolled interactions with the environment as well as crosstalk between system and measurement device lead to leakage of information out of the measured system. Several types of parity measurements have been implemented in circuit quantum electrodynamics, trapped ions and nuclear spins.  
Two-qubit parity measurements were realised non-destructively and repeatedly \citep{Sun2014,Cramer2015,Kelly2015,Negnevitsky2018}, and were used for demonstrations of multiple-round quantum error correction \citep{Cramer2015,Kelly2015},
to test quantum contextuality \citep{Kirchmair2009}, and for the preparation and stabilization of entangled states \citep{Riste2013,Pfaff2013, Saira2014,Chow2014,Sun2014,Cramer2015,Kelly2015,Negnevitsky2018}. 
Multi-qubit parity measurements have so far been limited to either destructive measurements as a benchmark for quantum processors \citep{Takita2016}, or to a single non-destructive measurement used to generate three-qubit \citep{Cramer2015} and four-qubit entangled states \citep{Barreiro2011}.
The ability to sequentially and non-destructively apply multi-qubit parity measurements would open up new opportunities for quantum error detection and -correction codes, state preparation and fundamental tests. 

In this manuscript, we experimentally realize repeated three-qubit parity measurements on nuclear spin qubits in diamond while minimizing the disturbance of the state of the qubits. We exploit these non-destructive measurements to deterministically generate a three-qubit Greenberger-Horne-Zeilinger (GHZ) state from any input state with three consecutive parity measurements. Finally, we demonstrate the usefulness of these measurements for fundamental tests by performing two contextuality experiments.

\begin{figure*}[t]%
\centering
\includegraphics[width=\textwidth]{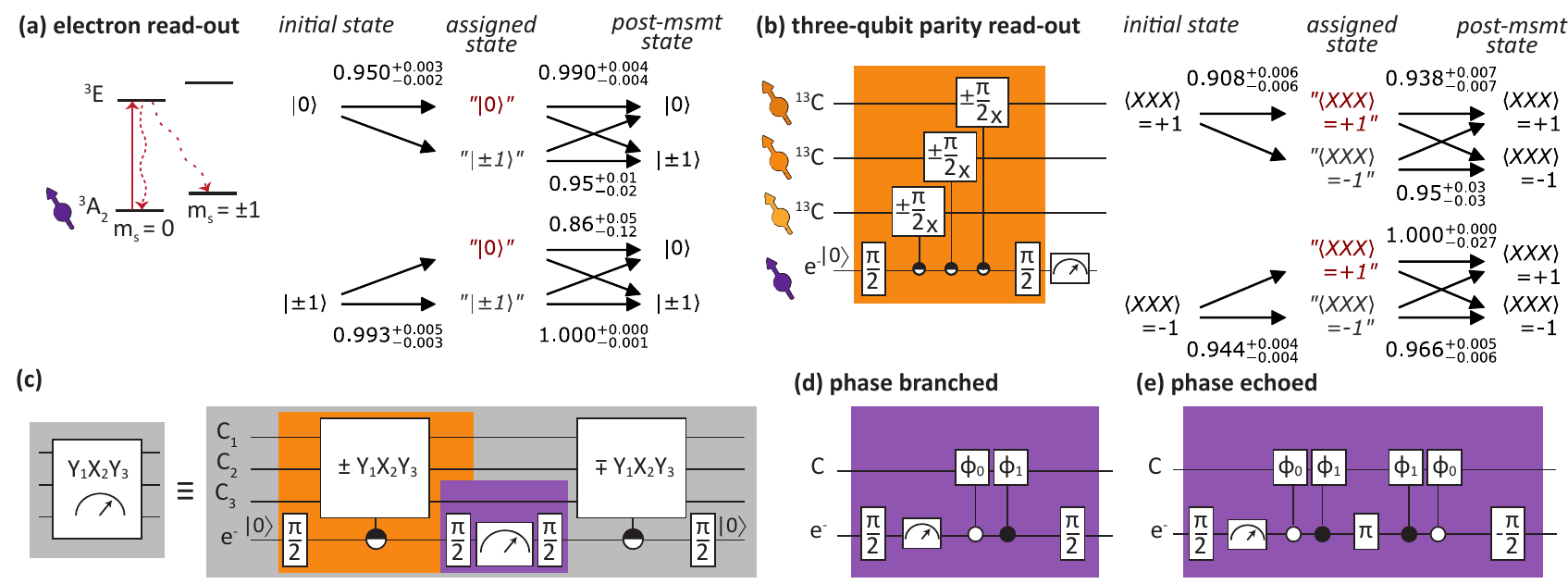}%
\caption{{\bf Three-qubit quantum parity measurements.} (color online) \textbf{(a)} The NV electron spin is read out in a single shot using spin-selective optical excitation. The readout fidelity and projectiveness are obtained with a maximum-likelihood estimation on raw data from repeated readouts \citep{suppl}. The data is analyzed assuming an electron spin initialisation fidelity of $0.998(2)$ and $0.995(5)$ for the $m_s=0$ and $m_s=\pm 1$ spin states respectively \citep{Robledo2011a}. \textbf{(b)} The joint parity state of three $^{13}$C nuclear spins is read out using the electron as ancilla. The positive (negative) parity subspace $\langle XXX\rangle~(-\langle XXX\rangle)$ is mapped onto electron state $|m_s=0\rangle~(|m_s=-1\rangle)$. 
The readout fidelity and projectiveness are obtained with a maximum-likelihood estimation on raw data from three repeated readouts on an initially mixed state. The model from which the fidelities are extracted does not take into account coherences or correlated errors between the measurements \citep{suppl}.
\textbf{(c)} After the parity readout an inverse unitary operation is applied to undo additional $\pi/2$ rotations on the nuclear spins and bring the electron ancilla back to $|0\rangle$, as needed to perform a next parity measurement. \textbf{(d)} Nuclear phases acquired when projecting on $m_s=0$ ($\phi_C = \phi_0$) or $m_s=- 1$ ($\phi_C = \phi_1$) during the readout can be separately tracked and included in the control sequence using fast-feedback \citep{Cramer2015} (nuclear phase accumulation is explicitly shown as conditional-phase gates). \textbf{(e)} Using an echo on the electron spin state, the nuclear states acquire a phase $\phi_C = \phi_0 + \phi_1$, that is independent of the measurement outcome, removing the need to branch the control sequence after each measurement.}
\label{fig:fig1}%
\end{figure*}

We implement the parity measurements on $^{13}$C nuclear spins in diamond that are weakly coupled via hyperfine interaction to the electron spin of a nitrogen-vacancy (NV) centre. These nuclear spins are an excellent workhorse for multi-qubit protocols \citep{Taminiau2014, Waldherr2014, Cramer2015} thanks to their long coherence times and their insensitivity to the optical and microwave fields that are used to control the NV centre electron. We use conditional quantum logic to map the parity of multiple nuclear spin states onto the electron spin that acts as an ancilla qubit. The electron is then read out in a single shot \citep{Robledo2011,Blok2014} (see \cref{fig:fig1}). In this way only the parity of the nuclear spin is projected and no information about the individual state of the nuclei is extracted, ensuring the non-destructive nature of the measurement. 

The measurement of the electron spin state is performed by optical excitation of a spin-dependent transition and detection of emitted photons. The cycling nature of the transition \citep{Tamarat2008,Robledo2011a} allows for a high readout fidelity, even for a finite photon detection efficiency. The readout fidelity and the non-destructive nature of the readout are limited by spin-flips during the readout. To maximize non-destructiveness we stop the optical excitation as soon as a photon is detected \citep{Blok2014, Cramer2015}. The resulting characterisation of assignment fidelity (the probability that the readout yields the correct outcome) and projectiveness (the probability that the state after the measurement corresponds to the assigned state) is shown in \cref{fig:fig1}a. From the diagram in \cref{fig:fig1}a we find the probability that the post-measurement state is the same as the initial state \citep{Lupascu2008,Riste2012}: 0.943(4) for $m_s=0$ and 0.991(3) for $m_s=\pm 1$.

The nuclear spin state is mapped onto the electron spin using electron-controlled nuclear spin rotations \citep{Taminiau2014} (\cref{fig:fig1}b). We apply sequences of electron $\pi$-pulses with an inter-pulse delay that is tuned to the hyperfine coupling of one of the weakly coupled nuclear spins to induce a rotation, while dynamically decoupling the electron state from the rest of the nuclear spin bath \citep{Taminiau2014}. Because the precession frequency of the nuclei depends on the electron spin state, the nuclear phases need to be carefully tracked throughout the experiment, based on knowledge of the electron spin state. An electron spin flip at an unknown time during the readout consequently dephases the nuclear state \citep{Blok2015,Reiserer2016,Kalb2018}. We find that this is one of the main sources of disturbance for the nuclear spin state during the parity measurement (see \cref{fig:fig1}b). Additional disturbances are due to imperfections of the electron-controlled gate. The probability that the system is in the same parity subspace before and after the measurement is 0.857(9) for the positive parity subspace, and 0.912(7) for the negative parity subspace. We note that this parity preservation by itself does not guarantee preservation of the coherences within the parity subspace.

Because the implementation of the electron-controlled gate deviates from a CNOT gate, the nuclear spins undergo an extra $\pi/2$ rotation (Fig. S2 of the Supplemental Material \citep{suppl}). To enable consecutive measurements along well-defined axes, we reverse the unitary operations that were used to map the nuclear spin parity onto the ancilla, in a way that is independent of the measurement outcome (see \cref{fig:fig1}c). This also resets the electron spin state to the initial state $m_s=0$. The sequence is compiled where possible: we remove unnecessary gates and adapt the gates based on the phase accumulated by the nuclear spins \citep{suppl}. 

The phase evolution of the nuclear spins depends on the electron spin readout outcome. A solution is to use fast-feedback to switch to a different branch of the control sequence after each readout, and track the phase acquired for readout outcomes $m_s=0$ and $m_s=-1$ separately \citep{Cramer2015} (\cref{fig:fig1}d). 
However, this leads to an outcome-dependence of the control sequence that complicates the interpretation of contextuality experiments.
In addition, if each branch is pre-programmed, this leads to memory requirements that are exponential in the number of measurements.
To avoid branching of the control sequences, we implement a spin echo \citep{Hahn1950} after the readout, leading to a phase evolution of each nuclear spin that is independent of the electron readout outcome (see \cref{fig:fig1}e). This removes any measurement outcome-dependence of the remainder of the sequence. In addition, it reduces the memory required to store control sequences from exponential to linear in the number of readouts, which is important for more complex protocols with more subsequent readouts.

We now use these three-qubit parity measurements for the creation of a maximally entangled three-qubit GHZ state \citep{Greenberger1990}. Preservation of the coherences within the parity subspaces after the parity measurement is crucial here: generating a GHZ state with three consecutive parity measurements is only possible if they are highly non-destructive. We consecutively apply the following parity measurements:
\begin{align}\label{eq:GHZ_observables}
p_1 &= \sigma_{x,1} \otimes \sigma_{y,2} \otimes \sigma_{y,3};\nonumber\\
p_2 &= \sigma_{y,1} \otimes \sigma_{x,2} \otimes \sigma_{y,3};\\
p_3 &= \sigma_{y,1} \otimes \sigma_{y,2} \otimes \sigma_{x,3},\nonumber 
\end{align}
where $\sigma_{x,k}$ and $\sigma_{y,k}$ are the $x$- and $y$-Pauli matrices on the $k$-th qubit. These measurements ideally project any input state of the three nuclear spins onto one of the eight GHZ states (e.g. $1/\sqrt{2}(|000\rangle+|111\rangle)$), depending on the measurement outcomes. To demonstrate this, we prepare the nuclear spins in the maximally mixed state before each measurement round by using resonant lasers that induce electron spin-flips and thereby dephase the nuclear spin states. 

Each parity measurement contains four or five electron-controlled nuclear spin rotations, that each consist of around 40 electron $\pi$-pulses. The total measurement sequence for GHZ state generation and verification spans a total time of approximately 10 ms. The dephasing times of the nuclear spin states are of the same order ($T^*_{2}=9.9(2) ~\text{ms}, 11.2(3)~\text{ms}, 17.3(6)~\text{ms}$ for nuclear spins 1, 2, and 3 respectively). However, dephasing is suppressed by the quantum Zeno effect \citep{Misra1977,Kalb2016}: repeated measurements project the state, restricting its evolution.

\begin{figure}[t]%
\centering
\includegraphics[width=\columnwidth]{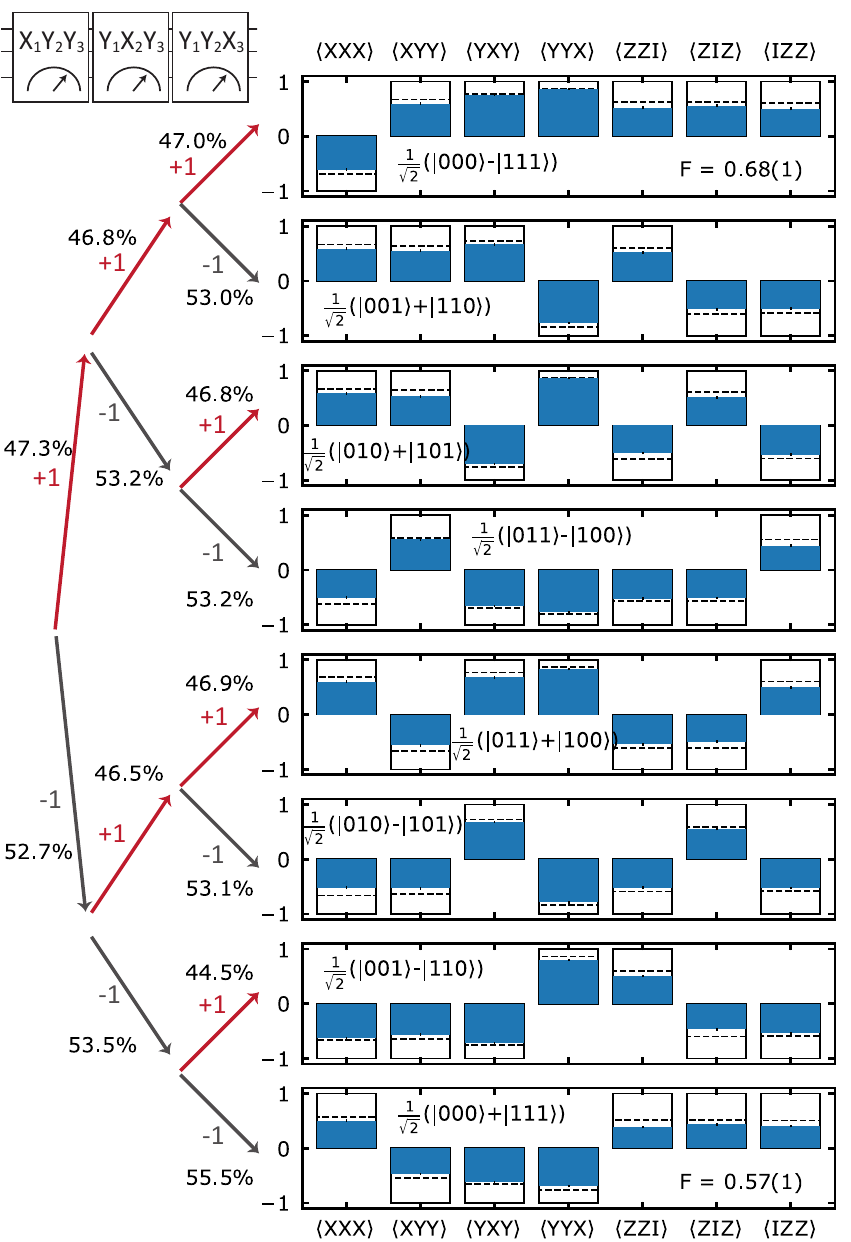}%
\caption{{\bf Creating a GHZ state by consecutive parity measurements.} (color online) Conditional on the outcomes of three consecutive parity measurements, one out of eight maximally entangled GHZ states is created out of an initially maximally mixed state. The control sequences used to obtain the data (blue filled bars) include phase branching after each parity measurement (\cref{fig:fig1}d). To better estimate the nuclear spin state, the final readout is corrected for electron spin readout infidelity. Positive parity is mapped onto the $m_s=0$ electron state during the measurements, such that the electron readout asymmetry results in the highest state fidelity for $\sqrt{1/2}(|000\rangle-|111\rangle)$. Black lines indicate the ideal outcome for a GHZ state, and black dashed lines are the outcome of a simulation with independently characterised parameters \citep{suppl}. Error bars (statistical standard error) on the data are indicated with black lines.}
\label{fig:fig2}%
\end{figure}

Measurements of the non-zero components of the resulting GHZ states on three nuclear spins are shown in \cref{fig:fig2}. 
As expected from the readout characterisation, we find that the best fidelity with a GHZ state ($F_\text{GHZ}=0.68(1)$) is obtained when positive parity (corresponding to the electron spin state $m_s=0$) is found three times in a row. But even when obtaining negative parity three times, the nuclear state has a fidelity $F_{\text{GHZ}}=0.57(1)$, still demonstrating genuine multipartite entanglement \citep{Acin2001,Bourennane2004}. The average fidelity for all eight states is 0.634(3). With these non-destructive parity measurements a multipartite entangled state can thus be deterministically prepared, as the long coherence times enable the application of feedback based on the measurement outcomes \citep{Cramer2015}. Importantly, these results show that the three-qubit parity measurements do not destroy coherences within the parity subspace.

Phase-echoed (\cref{fig:fig1}e) parity measurements (data is shown in the Supplemental Material \citep{suppl}) give an average GHZ state preparation fidelity of 0.600(3). We attribute the slight decrease in fidelity for the phase-echoed protocol to imperfect nuclear frequency calibration and imperfections in the electron echo pulse. We note that the good performance of the phase echoed implementation is promising as it enables extension to more complex protocols with more subsequent measurements, as required for e.g. quantum error correction. 

We next use sequential non-destructive parity measurements for a contextuality experiment. 
Quantum contextuality goes against the classical notion of noncontextuality: a measurement outcome should not depend on which other compatible measurements are performed jointly. Measurements are compatible if they can be measured jointly without disturbing each other's measurement outcome, i.e. observables A and B are compatible if measuring consecutively A-B-A gives twice the same outcome for A.

These classical versus quantum contradictions can be probed experimentally in a GHZ test \citep{Greenberger1990}.
In the original version of this test a system is prepared in a GHZ state and four sets of observables are measured: the three observables described in \cref{eq:GHZ_observables}, and a fourth observable,
\begin{align}\label{eq:GHZ_XXX_observable}
p_4 &= \sigma_{x,1} \otimes \sigma_{x,2} \otimes \sigma_{x,3}.
\end{align}
If we measure the first three sets of observables on the GHZ state $\sqrt{1/2}(|000\rangle+|111\rangle)$, we would, for ideal measurements, get the outcomes $(P_1,P_2,P_3)=(+1,+1,+1)$, where $P_j$ is the measurement outcome corresponding to measurement $p_j$. Given these three outcomes, a noncontextual theory predicts $P_4 = +1$. But quantum theory predicts $P_4 = -1$, thus showing a maximal contradiction with noncontextual models \citep{Greenberger1990}. 

In previous experiments the measurements $p_j$ ($j=1,...,4$)
were implemented as classical parity measurements: each qubit is measured individually and the parity calculated using the classical outcomes. Because these measurements do not preserve coherences between the qubits, each measurement $p_j$ needs to be performed separately on newly prepared GHZ states.
In that case the result can be formalized into an inequality as done by Mermin \citep{Mermin1990a}. This GHZ experiment can probe quantum nonlocality, and has been implemented in local \citep{Laflamme1998,Dicarlo2010,Neeley2010,Lanyon2014} and distant setups \citep{Pan2000a,Erven2014}. 

\begin{figure}[t]%
\centering
\includegraphics[width=\columnwidth]{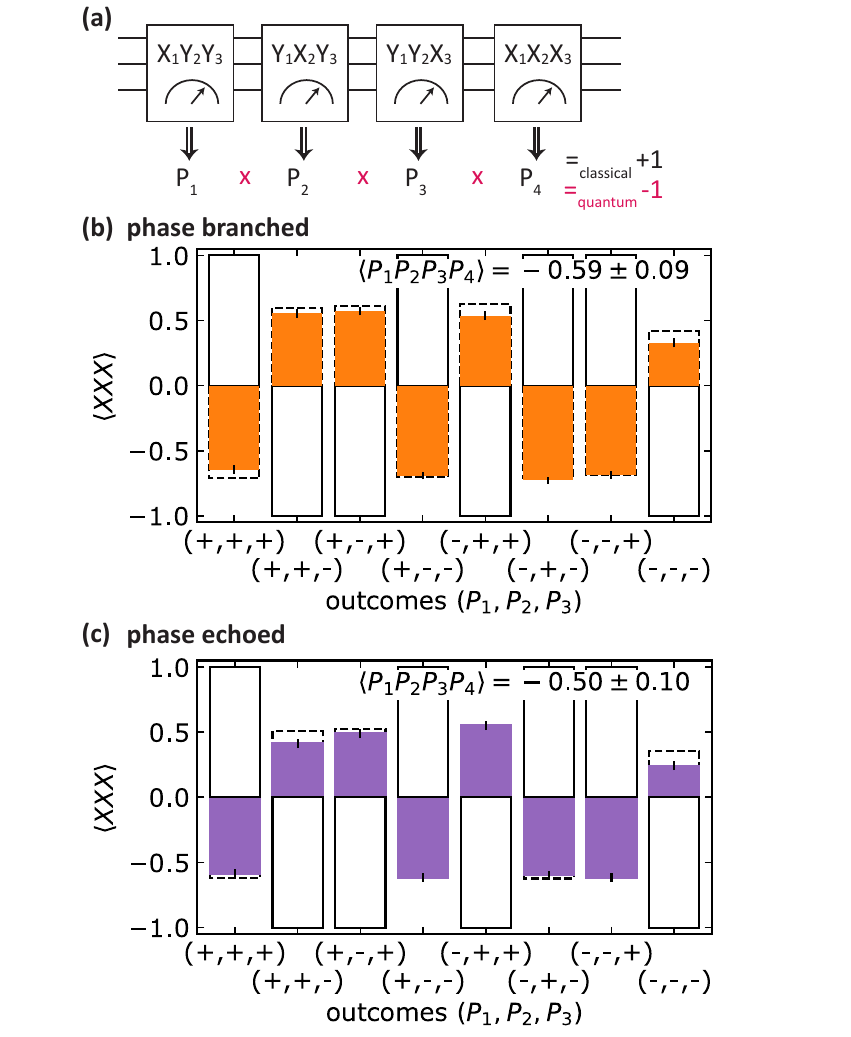}%
\caption{{\bf A single-shot GHZ experiment.} (color online) \textbf{(a)} For the consecutive application of four parity measurements a quantum versus classical contradiction is predicted in each measurement run. \textbf{(b-c)} Both the phase-branched (b, orange bars) and phase-echoed implementations (c, purple bars) of the protocol are in contrast with the classical expectation. Black lines indicate the classically expected outcomes, black dashed lines are the expected result from simulations with independently characterised parameters \citep{suppl}. Error bars (statistical standard error) on the data are indicated with black lines.}
\label{fig:fig3}%
\end{figure}

An interesting variation of the GHZ experiment has been proposed in which the measurements $p_j$ are performed as sequential quantum parity measurements on a single input state \citep{Lloyd1998,Laloe2012}. In such an implementation, a maximal quantum versus classical contradiction is obtained in every measurement round, since a non-contextual theory predicts $\langle P_1 \times P_2 \times P_3 \times P_4 \rangle = 1$, while quantum theory gives $\langle P_1\times P_2\times P_3 \times P_4 \rangle = -1$ (see \cref{fig:fig3}a). Imperfections in measurement assignment fidelity reduce the expectation value of the product but, for compatible measurements, cannot cause a sign flip. This single-shot form of the GHZ test is state-independent: the input of the measurement sequence does not have to be a maximally entangled GHZ state, but can be any state, even a mixed state. Such state-independence is a distinct feature of quantum contextuality tests \citep{Amselem2009,Kirchmair2009,Moussa2010,Zu2012,Huang2013,DAmbrosio2013,Zhang2013,Canas2014,Canas2014a,Leupold2018}.

We realize this single-shot GHZ experiment using parity measurements on nuclear spins both with the conventional phase-branched (\cref{fig:fig1}d) and new phase-echoed readout methods (\cref{fig:fig1}e). We find results contrasting the classically expected outcomes with both readout methods: $\langle P_1 \times P_2 \times P_3 \times P_4 \rangle = -0.58(6)$ and $-0.5(1)$ respectively, as shown in \cref{fig:fig3}b,c. 

\begin{figure}[t]%
\centering
\includegraphics[width=\columnwidth]{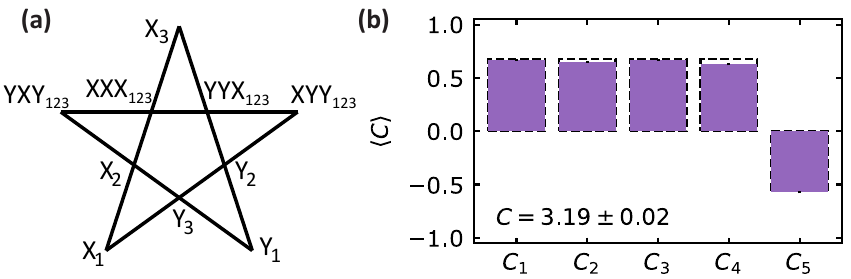}%
\caption{{\bf A noncontextuality test in 8 dimensions.} (color online) \textbf{(a)} A schematic representation of the five contexts in the noncontextuality inequality (NCI) as described in \cref{eq:C_ineq}. The vertices of the star represent a measurement; measurements along each side are compatible and form a context; noncontextual hidden variables assigned to each variable would satisfy $C\leq3$.  \textbf{(b)} Implementation of the NCI on nuclear spins in diamond using a phase-echoed (\cref{fig:fig1}e) readout for outcome-independence of the control sequence. The data shows a violation of the inequality bound ($C\leq 3$). Black dashed lines are the expected results from simulations \citep{suppl}; error bars (statistical standard error) on the data are indicated with black lines.
}
\label{fig:fig4}%
\end{figure}

The single-shot version of the GHZ experiment assumes that the parity measurements probe the individual underlying observables, e.g. that $\sigma_{x,1} \otimes \sigma_{y,2} \otimes \sigma_{y,3}$ probes $\sigma_{x,1}$, $\sigma_{y,2}$, and $\sigma_{y,3}$. 
The experiment can be extended to explicitly measure this, and to formalise the result through violation of an inequality. 
This is done by measuring four additional contexts (\cref{fig:fig4}a), testing a noncontextuality inequality (NCI) as proposed by Cabello \citep{Cabello2008}:
\begin{align} \label{eq:C_ineq}
&C=\langle C_1\rangle + \langle C_2\rangle + \langle C_3\rangle + \langle C_4\rangle - \langle C_5\rangle \leq 3; \\
&C_1 = X_1 \times Y_2 \times Y_3 \times P_1; \nonumber\\
&C_2 = Y_1 \times X_2 \times Y_3 \times P_2; \nonumber\\
&C_3 = Y_1 \times Y_2 \times X_3 \times P_3; \nonumber\\
&C_4 = X_1 \times X_2 \times X_3 \times P_4; \nonumber\\
&C_5 = P_1 \times P_2 \times P_3  \times P_4, \nonumber
\end{align}
where $X_k$ and $Y_k$ are the measurement outcomes corresponding to $\sigma_{x,k}$ and $\sigma_{y,k}$.
The bound of 3 is found for noncontextual hidden-variable (NCHV) models, while for an ideal quantum system with perfectly non-destructive quantum parity measurements, $C=5$ is predicted.
Like the single-shot GHZ experiment, this NCI is state-independent.
To test it requires the application of up to four consecutive three-qubit parity measurements in an eight-dimensional system. So far, the highest-dimensional state-independent NCI that has been tested featured three sequential two-qubit parity measurements in a four-dimensional system \citep{Kirchmair2009}.

We implement the NCI using the phase-echoed nuclear spin parity measurements (see \cref{fig:fig4}b) and observe a violation of the noncontextual bound: $C = 3.19(2)$, rejecting the hypothesis that our experiment is described by a non-contextual model with a p-value of $1.21\times 10^{-14}$ \citep{Elkouss2016, suppl}.
Note that, as in any such contextuality test, the measurements must be assumed to be compatible in order to reach this conclusion \citep{Guhne2010}. Since we efficiently detect the observables, no fair sampling assumption is necessary.
With improved experimental parameters, e.g. using refocusing pulses on the nuclear spin states, decoherence-protected subspaces \citep{Reiserer2016} or isotopic purification of the nuclear environment \citep{Bar-Gill2013}, experiments may be designed in which theories that incluse specific models for measurement incompatibility can be further restricted \citep{Guhne2010}.

In conclusion, we realized repeated non-destructive three-qubit parity measurements. We use a readout echo pulse to prevent nuclear phase branching, enabling memory-efficient and outcome-independent implementation of sequential measurements. We apply three-qubit parity measurements on a maximally mixed state to generate genuine multipartite entanglement, and we realize a test of quantum contextuality in a single-shot. Furthermore, we push the implementation of noncontextuality tests to higher-dimensional systems than previously reported. The techniques and insights developed here can be directly applied to parity-measurement-based quantum computing protocols such as quantum error correction \citep{Bravyi1998,Raussendorf2007,Cramer2015}.

\begin{acknowledgements}
We thank M. Abobeih, C. Bradley, N. Bultink, A. Cabello, L. DiCarlo, K. Goodenough, P.C. Humphreys, N. Kalb, and M.A. Rol for helpful discussions. We acknowledge support from the Netherlands Organisation for Scientific Research (NWO) through a VICI grant and a VIDI grant, the European Research Council through a Synergy Grant, and the Royal Netherlands Academy of Arts and Sciences and Ammodo through an Ammodo KNAW Award. J.C. acknowledges support from the NWO Graduate Programme.
\end{acknowledgements}

\end{document}